\documentclass[onecolumn,aps,prl,superscriptaddress]{revtex4}
\usepackage{mathrsfs}
\usepackage{graphicx}
\usepackage{amsfonts}
\usepackage{amsmath}
\usepackage{amssymb}

\begin{document}


\title{All-Versus-Nothing Proof of Einstein-Podolsky-Rosen Steering}



\author{Jing-Ling~Chen\footnote{Correspondence and requests for materials should be addressed to
J.L.C. (cqtchenj@nus.edu.sg).}}
 \affiliation{Theoretical Physics Division, Chern Institute of Mathematics, Nankai University,
 Tianjin 300071, People's Republic of China}
 \affiliation{Centre for Quantum Technologies, National University of Singapore,
 3 Science Drive 2, Singapore 117543}

\author{Xiang-Jun~Ye}
 \affiliation{Theoretical Physics Division, Chern Institute of Mathematics, Nankai University,
 Tianjin 300071, People's Republic of China}
 \affiliation{Centre for Quantum Technologies, National University of Singapore,
 3 Science Drive 2, Singapore 117543}

\author{Chunfeng~Wu}
 \affiliation{Centre for Quantum Technologies, National University of Singapore,
 3 Science Drive 2, Singapore 117543}

\author{Hong-Yi~Su}
 \affiliation{Theoretical Physics Division, Chern Institute of Mathematics, Nankai University,
 Tianjin 300071, People's Republic of China}
 \affiliation{Centre for Quantum Technologies, National University of Singapore,
 3 Science Drive 2, Singapore 117543}

\author{Ad\'{a}n~Cabello}
 \affiliation{Departamento de F\'{\i}sica Aplicada II, Universidad de
 Sevilla, E-41012 Sevilla, Spain}

\author{L.~C.~Kwek}
 \affiliation{Centre for Quantum Technologies, National University of Singapore,
 3 Science Drive 2, Singapore 117543}
 \affiliation{National Institute of Education and Institute of Advanced Studies,
 Nanyang Technological University, 1 Nanyang Walk, Singapore 637616}

\author{C.~H.~Oh}
 \affiliation{Centre for Quantum
Technologies, National University of Singapore, 3 Science Drive 2,
Singapore 117543}
\affiliation{Department of Physics, National
University of Singapore, 2 Science Drive 3, Singapore 117542}

\date{\today}
\maketitle


\textbf{Einstein-Podolsky-Rosen steering is a form of quantum
nonlocality intermediate between entanglement and Bell nonlocality.
Although Schr\"odinger already mooted the idea in 1935, steering
still defies a complete understanding. In analogy to
``all-versus-nothing'' proofs of Bell nonlocality, here we present a
proof of steering without inequalities rendering the detection of
correlations leading to a violation of steering inequalities
unnecessary. We show that, given any two-qubit entangled state, the
existence of certain projective measurement by Alice so that Bob's
normalized conditional states can be regarded as two different pure
states provides a criterion for Alice-to-Bob steerability. A
steering inequality equivalent to the all-versus-nothing proof is
also obtained. Our result clearly demonstrates that there exist many
quantum states which do not violate any previously known steering
inequality but are indeed steerable. Our method offers advantages
over the existing methods for experimentally testing steerability,
and sheds new light on the asymmetric steering problem.}




\vspace{8pt}

Quantum nonlocality is an invaluable resource in numerous quantum information
protocols. It is part of a hierarchical structure \cite{WJD07}:
quantum states that have Bell nonlocality \cite{Bell} form a subset
of Einstein-Podolsky-Rosen steerable states which, in turn, form a
subset of entangled states. The concept of steering can historically be traced
back to Schr\"odinger's reply~\cite{Schrodinger35} to the
Einstein-Podolsky-Rosen argument~\cite{EPR}, and it has since been rigorously
formulated by Wiseman, Jones, and Doherty \cite{WJD07}.

Within the steering scenario, Alice prepares a bipartite system, keeps
one particle and sends the other one to Bob. She announces that the
Bob's particle is entangled with hers, and thus
that she has the ability to ``steer'' the state of Bob's particle at
a distance. This means that she could prepare Bob's particle in
different states by measuring her particle using different settings.
However, Bob does not trust Alice; Bob worries that she may send him
some unentangled particles and fabricate the results using her
knowledge about the local hidden state (LHS) of his particles. Bob's
task is to prove that no such hidden states exist.

The study of Bell nonlocality have witnessed phenomenal developments to
date with important widespread applications \cite{Ekert,Brukner,Random10}.
Its existence can be demonstrated through two different approaches:
the first concerns the violations of Bell inequalities, and the second relies on an
all-versus-nothing (AVN) proof without inequalities
\cite{GHZ89,Hardy,Cabello01-1,Cabello01-2}. The AVN proof shows a
logical contradiction between the local-hidden-variable models and
quantum mechanics, and thus offers an elegant argument of the
nonexistence of local-hidden-variable models. What is possible with
Bell nonlocality and local hidden variables
should also be possible with steering and local hidden states.
In stark contrast to Bell nonlocality, the study of steering is still at its infancy.
Recent works like Refs.~\cite{WJD07, steering2} put steering on firmer grounds.
Like Bell nonlocality, this topic is generally of broad interest, as it
hinges on questions pertaining to the foundations of quantum
physics~\cite{SU}, and at the same time reveals new possibilities
for quantum information~\cite{QKD}. Einstein-Podolsky-Rosen steering
can be detected through the violation of a steering inequality,
which rules out the LHS model in the same spirit in which the
violation of a Bell inequality rules out the local-hidden-variable
model. Recently, several steering inequalities have been proposed
and experimentally tested
\cite{NP2010,SGDBFWLCGWNW12,BESBCWP11,WRSLBWUZ11}.
Nevertheless,
steering is far from being completely understood and the subject
deserves further investigation.

The AVN proof for Bell nonlocality
\cite{GHZ89,Hardy,Cabello01-1,Cabello01-2} has been developed to
rule out any local-hidden-variable models. Likewise, it is
interesting to find out if there an analogous AVN proof which
can rule out any LHS models for steering. The purpose of this work is
to present an affirmative answer to this question by showing that
Einstein-Podolsky-Rosen steering without inequalities exists in a
two-qubit system. This proof is an analogy of AVN argument for Bell
nonlocality without inequalities, and offers advantages over the
existing methods for experimentally testing steerability as well as shedding
new light on the asymmetric steering problem. In addition, a
steering inequality based on the AVN proof is
also obtained.


\vspace{12pt} \noindent{\bf Results} \vspace{8pt}


\begin{figure}[tbp]
\includegraphics[width=150mm]{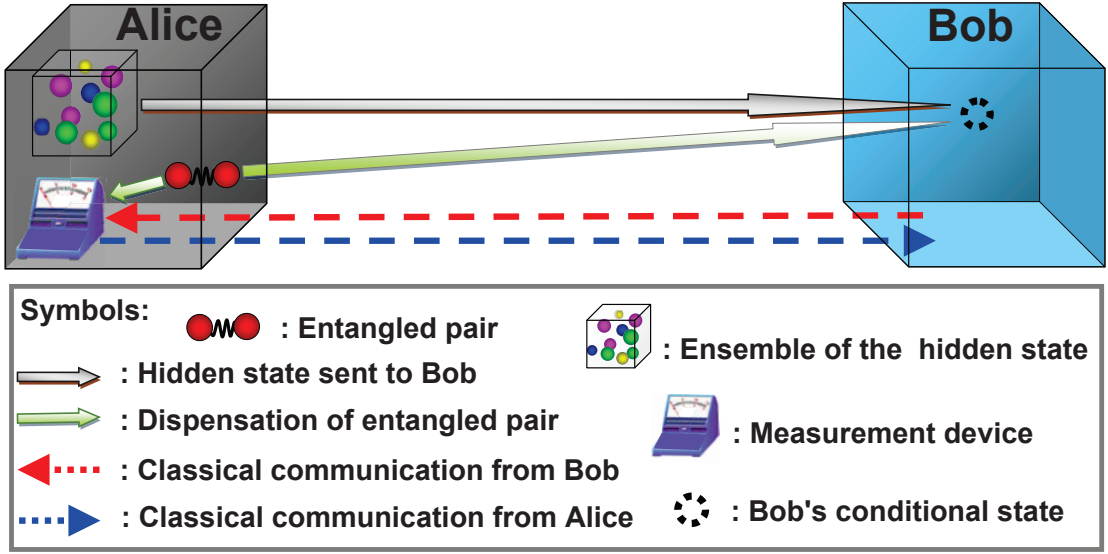}\\
\caption{\textbf{The steering scenario illustration.} Alice first
prepares a two-qubit state and keeps
one qubit. She then sends the other qubit to Bob and announces that it is entangled with
the one she possesses (see the pair of red balls and green arrows). Thus
she could remotely ``steer'' Bob's state by projective measurements. However, Bob does not trust Alice and he worries that she may fabricate the results using her knowledge about LHS.
In the two-setting steering
scenario, Bob asks Alice to perform two specific projective measurements
on her qubit (see the red dashed arrow) and
to let him know the measurement results (see the blue dashed
arrow). After Alice's measurement (see the measurement device), Bob
obtains four conditional states (see
the dashed circle). Alice could cheat Bob if there exists an ensemble
(see the gray box with colored
particles) and a stochastic map, such that the set of equations (\ref{E0}) holds.
To be convinced that
Alice can steer his state, Bob needs to confirm that no such hidden
states are possible.
 }\label{AliceBob}
\end{figure}


\noindent{\textbf{Steering without inequalities for two qubits.}}
The two-setting steering scenario can be described as follows: at the
beginning, Alice prepares a two-qubit state $\rho_{AB}$. She keeps
one qubit and sends the other to Bob. She then announces that it is entangled with
the one she holds (see Fig. \ref{AliceBob}), and that
she could remotely ``steer'' his state by projective measurements
$\mathcal {P}^{\hat{n}}_a=[\openone+(-1)^a {\hat{n}}\cdot {\vec
\sigma}]/2$, with $\hat{n}$ the measurement direction, $a$ (with
$a=0,1$) the Alice's measurement result, $\openone$ the $2 \times 2$
identity matrix, and ${\vec \sigma}=(\sigma_x, \sigma_y, \sigma_z)$
the vector of the Pauli matrices. Bob then
asks Alice to perform two projective measurements
$\mathcal {P}^{\hat{n}_1}_a$ and $\mathcal {P}^{\hat{n}_2}_a$ (with
$\hat{n}_1\neq \hat{n}_2$) on her qubit and
to tell him the measurement results of $a$.
After Alice's measurement has been done, Bob
obtains the four conditional states $\tilde{\rho}^{\hat{n}_j}_a$.
Alice could cheat Bob if there exists an ensemble
$\{ \wp_{\xi} \rho_{\xi} \}$ (see the gray box with colored
particles in Fig. \ref{AliceBob}) and a stochastic map $\wp(a|\hat{A},\xi)$ from $\xi$ to
$a$, such that the following equations 
hold,
\begin{eqnarray}
 \tilde{\rho}^{\hat{n}_j}_a&=& \sum_\xi \wp(a|\hat{n}_j,\xi) \wp_{\xi} \rho_{\xi},\;\;(a=0,1; j=1,2).\label{E0}
\end{eqnarray}
In order for Bob to be convinced that
Alice can steer his state, Bob needs to be sure that no such hidden
states are indeed possible.
If we demand that Bob's states possess an LHS description, then
his density matrices should satisfy Eq.~(\ref{E0}).
A contradiction among the four equations, meaning that they cannot
have a common solution of $\{ \wp_{\xi} \rho_{\xi} \}$ and
$\wp(a|\hat{n},\xi)$), convinces Bob that an LHS model does
not exist and that Alice can steer the state of his qubit.

It is worth mentioning that the set of equations (\ref{E0}) plays an
analogous role to the one in the standard
Greenberger-Horne-Zeilinger (GHZ) argument \cite{GHZ89}. The principal
difference between the arguments is that the set of equations in (\ref{E0}) deal with density matrices whereas in the GHZ argument, each equation pertains to the outcomes of measurements and therefore corresponds to real numbers.
The constraints imposed by LHS model on density matrices are much stricter than constraints imposed by real numbers. This provides an intuitive explanation as to why AVN proof would work for the Einstein-Podolsky-Rosen steering of two-qubit states.

Suppose that Alice initially prepares a product state
$\rho_{AB}=|\psi_A\rangle\langle\psi_A|\otimes
|\psi_B\rangle\langle\psi_B|$. It can be verified that, for any
projective measurement $\mathcal {P}^{\hat{n}}_a$ (with $\mathcal
{P}^{\hat{n}}_a \neq |\psi_A\rangle\langle\psi_A|$ and
$|\psi^\perp_A\rangle\langle\psi_A^\perp|$) performed by Alice, Bob
always obtains two identical pure normalized conditional states as
$\rho^{\hat{n}}_a=\tilde{\rho}^{\hat{n}}_a/{\rm
tr}\tilde{\rho}^{\hat{n}}_a=|\psi_B\rangle\langle\psi_B|,\;(a=0,1)$,
which means that Alice cannot steer Bob's state. Moreover, Bob can
obtain two identical pure normalized conditional states if and only
if $\rho_{AB}$ is a direct-product state. Hence, hereafter we
assume that $\rho^{\hat{n}}_0$ and $\rho^{\hat{n}}_1$ are two
different pure states, i.e., $\rho^{\hat{n}}_0 \neq
\rho^{\hat{n}}_1$.


For a general $\rho_{AB}$, $\rho^{\hat{n}}_a$
 are not pure. If they are pure, then $\rho_{AB}$ possesses the following uniform form:
\begin{eqnarray}
 \rho_{AB}&=&\mathcal {P}^{\hat{n}}_0 \otimes \tilde{\rho}^{\hat{n}}_0+\mathcal {P}^{\hat{n}}_1 \otimes \tilde{\rho}^{\hat{n}}_1 +|+\hat{n}\rangle\langle -\hat{n}|\otimes \mathcal{M}\nonumber\\
 &&+|-\hat{n}\rangle\langle +\hat{n}|\otimes \mathcal {M}^\dagger,\nonumber
\end{eqnarray}
where $|\pm \hat{n}\rangle$ are eigenstates of $\hat{n}\cdot \vec
\sigma$, $\mathcal {M}$ is a $2\times 2$ complex matrix under the
positivity condition of $\rho_{AB}$, and $\mathcal {M}^\dagger$ is
the Hermitian conjugation of $\mathcal {M}$.

For $\rho_{AB}$, it is not difficult to find that
$\mathcal {M}=\textbf{0}$ if and only if $\rho_{AB}$ is separable, and
the state $\rho_{AB}$ admits a LHS (which means that it is not steerable) if and only if $\mathcal {M}=\textbf{0}$ (see the Methods section).
In a two-setting steering protocol of $\{\hat{n}_1, \hat{n}_2\}$, if Bob
can obtain two different pure normalized conditional states along
Alice's projective direction $\hat{n}_1$ (or $\hat{n}_2$),
 the following three propositions are equivalent:
(i) $\mathcal {M}\neq0$. (ii) $\rho_{AB}$ is entangled. (iii) No LHS model exists for Bob's states,
so $\rho_{AB}$ is steerable (in the sense of Alice steering Bob's state).
We thus have our steering argument concluded, and that is given any two-qubit entangled state, the
existence of certain projective measurement by Alice so that Bob's normalized conditional states are two
different pure states provides a criterion for Alice-to-Bob steerability.

Although the standard GHZ argument is elegant for providing a full
contradiction between local-hidden-variable model and quantum
mechanics (with $100\%$ success probability), its validity is only
limited to some pure states with high symmetry, such as $N$-qubit
GHZ states and cluster states with $N\ge 3$ \cite{GHZ-cluster}.
Hardy attempted to extend the GHZ argument to an arbitrary two-qubit
system~\cite{Hardy}. However, Hardy's argument works for only $9\%$
of the runs of a specially constructed experiment. Moreover, Hardy's
proof is not valid for two-qubit maximally entangled state. To
overcome this, Cabello proposed an AVN proof for two observers, each
possessing a two-qubit maximally entangled
state~\cite{Cabello01-1,Cabello01-2}. Nowadays, there is no AVN proof of Bell nonlocality for a genuine
two-qubit state presented. However, we show
that for any two-qubit entangled state $\rho_{AB}$, if there exists
a projective direction $\hat{n}$ such that Bob's normalized
conditional states $\rho^{\hat{n}}_a$ become two different pure
states, then Alice can steer Bob's state. Our steering argument is
not only valid for two-qubit pure states, but it is also applicable
to a wider class of states including mixed states.


\begin{figure}[tbp]
\includegraphics[width=150mm]{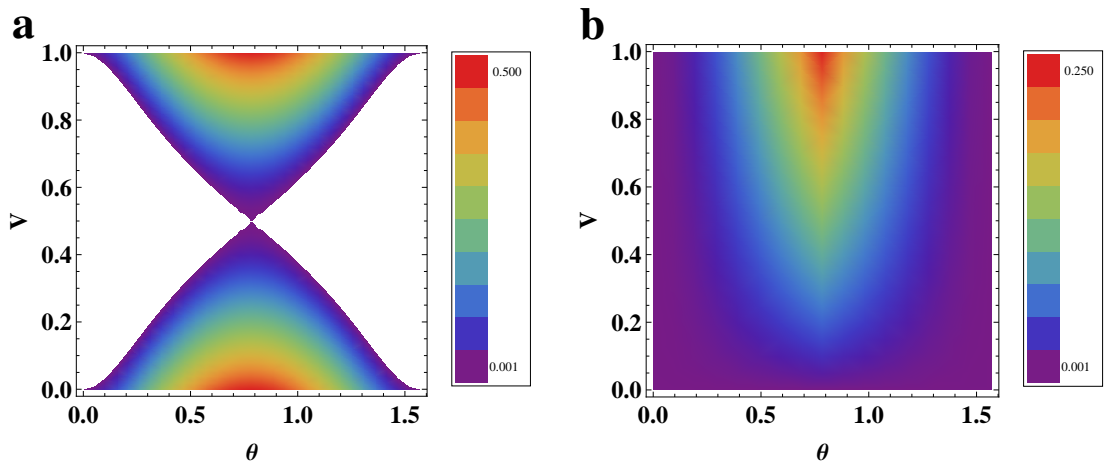}\\
\caption{\textbf{(a) Detecting steerability of the state
(\ref{teststate}) using the ten-setting steering inequalities.} We explore the steering of state (\ref{teststate}) via violation
of the ten-setting inequality presented in Ref.~\cite{NP2010}. The
colors denote different values of quantum violation, as scaled in the legend.
The blank region indicates that steerability of (\ref{teststate})
cannot be detected by this inequality. With the replacement
$A_k\rightarrow \vec{\sigma}_k^A$ and $\vec{\sigma}_k^B\rightarrow
B_k$ in the above inequality, one obtains a similar steering
inequality $\mathcal{S}'_{N} \leq 0$ to show Bob's ability of
steering Alice's state. The inequality $\mathcal{S}'_{N=10} \leq 0$
yields the same violation region. This indicates that steering
inequalities in Ref.~\cite{NP2010} cannot reveal asymmetric
steering. \textbf{(b) Detecting steerability using the steering
inequality (\ref{inequlitys}).}
We show the steering of the state $\rho_{\rm
col}$ through violation of inequality (\ref{inequlitys}).
Quantum prediction of the left-hand-side of the inequality always succeeds $0$ unless $V=0$ or
$\theta=0,\pi/2$.
 }\label{Fig1}
\end{figure}


\vspace{8pt} \noindent{\textbf{{The AVN proof versus the known
steering inequalities.}} Let us compare our result with the known
steering inequalities. First, they play different roles in
demonstrating steering: steering inequality follows a similar
approach to the Bell inequality for Bell nonlocality, while steering
without inequality serves as an analogous counterpart to the GHZ test of Bell
nonlocality without Bell inequalities.

Secondly, our argument shows that there are many quantum steerable states that
do not violate any known steering inequalities. For an example, consider the state
\begin{equation}
 \label{teststate}
 \rho_{V}^{\theta}=V |\Psi(\theta)\rangle\langle\Psi(\theta)|
 +(1-V)|\Phi(\theta)\rangle\langle\Phi(\theta)|,
\end{equation}
where $
|\Psi(\theta)\rangle=\cos\theta|00\rangle+\sin\theta|11\rangle$,
$|\Phi(\theta)\rangle=\cos\theta|10\rangle+\sin\theta|01\rangle$.
It is entangled when $V\in[0,1/2)\cup(1/2,1]$ and $\theta\in (0,
\pi/2)$. It can be easily verified that, for state (\ref{teststate}), after Alice performs an
$\hat{x}$-direction measurement on her qubit, Bob's normalized
conditional states are just two different pure states, $\cos\theta|0\rangle+\sin\theta|1\rangle$ and $\cos\theta|0\rangle-\sin\theta|1\rangle$. Thus, based on our AVN proof of steering, Alice can always steer Bob's state using just a two-setting
protocol $\{\hat{x}, \hat{z}\}$.
On the other hand, a class of $N$-setting steering inequality
$\mathcal{S}_{N}=\frac{1}{N}\sum_{k=1}^N\langle
A_k\vec{\sigma}_k^B\rangle-C_N \leq 0$ has been introduced in Ref.~\cite{NP2010} to show the ability of Alice steering Bob's state.
By running a numerical check of a 10-setting steering inequality of the above form, we observe that,
for some regions of $V$ and $\theta$, the steering inequality cannot detect the steering of state (\ref{teststate}) (as shown in Fig.~\ref{Fig1} \textbf{a}).
The colors denote different violation values, as shown in the legend.
The blank region indicates that the steerability of state (\ref{teststate})
cannot be detected by resorting to this inequality.

Finally, unlike quantum entanglement and Bell nonlocality, the
definition of steering is \emph{asymmetric}~\cite{WJD07, AS12}.
Our AVN proof can shed light on this problem. The state~(\ref{teststate})
is not symmetric under a permutation of Alice and Bob
(even with local unitary transformations acting on the state).
The known steering inequalities in Ref.~\cite{NP2010}
do not reveal asymmetric steering (see Fig.~2 \textbf{a}). However,
our argument presents a promising way to reveal
asymmetric steering. According to our AVN proof, the state
(\ref{teststate}) exhibits two-setting asymmetric steering.
On one hand, Alice can always steer Bob's state using just the two-setting
protocol $\{\hat{x}, \hat{z}\}$. On the other hand,
after Bob has performed a projective measurement along
an arbitrary $\hat{n}$-direction on his qubit, Alice's normalized
conditional states can never be cast into two different pure states, allowing
for the existence of LHS models. Take the state with parameters $V=3/5$ and
$\theta=\pi/8$ as an example (whose corresponding point is outside
of the colored region in Fig.~\ref{Fig1} \textbf{a}): Numerical results show
that, for any two-setting protocol $\{\hat{n}_1,\hat{n}_2\}$, there
is always a solution of LHS for Alice's conditional states. In short, this example illustrates
a state in which the steering scenario is not interchangeable.
This result can be of practical importance, since asymmetric steering
has applications in one-way quantum cryptography \cite{AS08} and may have potential applications in other fields of quantum information processing.


\vspace{8pt} \noindent{\textbf{A steering inequality.}} It is known that a Bell inequality can be derived
from the GHZ argument \cite{Mermin90}. This is also the case for the
steering without inequalities argument. The steering inequality
equivalent to the AVN proof reads
\begin{eqnarray}\label{inequlitys}
\langle \mathscr{W}_3\rangle - C_{\rm LHS}\leq 0,
\end{eqnarray}
subject to the constraint $\langle \mathscr{W}_1\rangle=\langle
\mathscr{W}_2\rangle=0$. Here $\mathscr{W}_j$ are projectors as
$\mathscr{W}_1=\mathcal {P}^{\hat{n}}_0 \otimes
{{\rho}^{\hat{n}}_0}^\perp, \mathscr{W}_2=\mathcal {P}^{\hat{n}}_1
\otimes {{\rho}^{\hat{n}}_1}^\perp, \mathscr{W}_3=|+\rangle\langle
+|\otimes |\hat{n}_B\rangle\langle \hat{n}_B|$, with
${{\rho}^{\hat{n}}_a}^\perp$ orthogonal to ${{\rho}^{\hat{n}}_a}$,
$|+\rangle=(|+\hat{n}\rangle+|-\hat{n}\rangle)/\sqrt{2}$,
$|\hat{n}_B\rangle=\cos\frac{\theta_B}{2}|0\rangle+\sin\frac{\theta_B}{2}e^{i\varphi_B}|1\rangle$,
$\langle \mathscr{W}_j\rangle= {\rm tr}( \mathscr{W}_j
\varrho_{AB})$, and $C_{\rm LHS}=\max_{\hat{n}_B}\left[{\rm
tr}\left(|\hat{n}_B\rangle\langle \hat{n}_B|
(\tilde{\rho}^{\hat{n}}_0+\tilde{\rho}^{\hat{n}}_1)/2\right)\right]$ is the upper bound for the LHS
model. Its physical implication can be described as
follows: Suppose Alice performs a
projective measurement in the $\hat{n}$-direction and finds that Bob
can obtain two different pure normalized conditional states, then
$\langle \mathscr{W}_1\rangle=\langle
\mathscr{W}_2\rangle=0$. They then perform a joint-measurement
$\mathscr{W}_3$ (in which Alice's measurement direction is
perpendicular to $\hat{n}$-direction). According to Lemma 2 (see the Methods section), the LHS
model requires $\mathcal {M}=0$, thus the probability $\langle
\mathscr{W}_3\rangle$ is bounded by $C_{\rm LHS}$. However, with quantum
mechanics, this bound is always exceeded due to a non-vanishing
$\mathcal{M}$.

As an instance, we investigate the steering of state $\rho_{\rm
col}=V|\Psi(\theta)\rangle\langle \Psi(\theta)|+(1-V)\openone_{\rm
col}$, with color noise $\openone_{\rm col}=(|00\rangle\langle
00|+|11\rangle\langle 11|)/2$ by using our inequality (\ref{inequlitys}). We find that Bob's conditional states
on Alice's projective measurement in the $z$-direction are two
different pure states $|0\rangle\langle 0|$ and $|1\rangle\langle
1|$, and the upper bound is $C_{\rm LHS}=(1+V|\cos2\theta|)/4$.
The quantum prediction of the left-hand-side of inequality (\ref{inequlitys}) reads $\frac{1}{2}V\sin^2\theta$ for
$\theta\in[0,\pi/4]$, and $\frac{1}{2}V\cos^2\theta$ for
$\theta\in[\pi/4,\pi/2]$, which do not vanish unless $V=0$ or
$\theta=0,\pi/2$ (see Fig.~2 \textbf{b}). The violation of the inequality clearly demonstrates that
the state $\rho_{\rm col}$ possesses steerability except $V=0$ or
$\theta=0,\pi/2$.


\vspace{12pt} \noindent{\bf Discussions} \vspace{8pt}


\noindent We have presented an AVN proof of Einstein-Podolsky-Rosen
steering for two qubits without inequalities based on a two-setting
steering protocol. The argument is valid for any two-qubit entangled
state, both pure and mixed. We show that many quantum states that
do not violate any known steering inequalities are indeed
steerable states. This provides a new perspective for understanding
steerability and offers an elegant argument for the nonexistence of LHS
models without resorting to steering inequalities. The result also sheds new
light on the asymmetric steerability -- a phenomenon with no counterpart
in quantum entanglement and Bell nonlocality. The result is testable
through measurements of Bob's conditional states and provides a simple
alternative to the existing experimental method for detecting
steerability \cite{NP2010,SGDBFWLCGWNW12,BESBCWP11,WRSLBWUZ11}.
Theoretically, a two-setting steering protocol can be used to show
that no LHS models exist for $\rho_{AB}$ if the state satisfies the
condition given in our AVN argument. Experimentally, the
determination of the steerability of a quantum state can be done by
performing quantum state tomography~\cite{James} on Bob's qubit.
Moreover, a steering inequality is obtained from our AVN argument,
and this inequality offers another way to test steerability of states.
Like Bell nonlocality whose importance has only
been realized with the rapid development of
quantum information science, we anticipate
further developments in this exciting area.


\vspace{12pt} \noindent{\bf Methods} \vspace{8pt}


We prove two Lemmas in the section.
The steerability of $\rho_{AB}$ is equivalent to that of the state
$\varrho_{AB}=(\mathcal {U}_A\otimes \openone)\rho_{AB}(\mathcal
{U}^\dagger_A\otimes \openone)$. It is always possible for Alice to
choose an appropriate unitary matrix $\mathcal {U}$ that rotates the
direction $\hat{n}$ to the direction $\hat{z}$. Therefore, we can
initially set $\hat{n}=\hat{z}$ by studying the state $\varrho_{AB}$
instead of $\rho_{AB}$. After Alice performs a projective
measurement in the $\hat{z}$-direction, Bob's unnormalized
conditional states are
\begin{subequations}\label{z01}
\begin{eqnarray}
 \tilde{\rho}^{\hat{z}}_0={\rm tr}_A[(|0\rangle\langle 0|\otimes \openone)\varrho_{AB}]=\mu_1
 |\varphi_1\rangle\langle\varphi_1|,\label{z01a}\\
 \tilde{\rho}^{\hat{z}}_1={\rm tr}_A[(|1\rangle\langle 1|\otimes \openone)\varrho_{AB}]=\mu_2
 |\varphi_2\rangle\langle\varphi_2|,\label{z01b}
\end{eqnarray}
\end{subequations}
with $\mu_1={\rm tr}(\tilde{\rho}^{\hat{z}}_0)$, $\mu_2={\rm
tr}(\tilde{\rho}^{\hat{z}}_1)$, ${\rho}^{\hat{z}}_0=
|\varphi_1\rangle\langle\varphi_1|$, and
${\rho}^{\hat{z}}_1=|\varphi_2\rangle\langle\varphi_2|$. Then one
has
\begin{eqnarray}
 \varrho_{AB}&=&\mu_1|0\rangle\langle 0|\otimes |\varphi_1\rangle\langle \varphi_1|+\mu_2|1\rangle\langle 1|\otimes|\varphi_2\rangle\langle \varphi_2|\nonumber\\
&&+|0\rangle\langle 1|\otimes \mathcal {M}+|1\rangle\langle
0|\otimes \mathcal {M}^\dagger. \label{varrhoAB}\nonumber
\end{eqnarray}


\noindent \emph{Lemma 1:} $\mathcal {M}=\textbf{0}$ if and only if $\varrho_{AB}$ is separable.


\noindent\emph{Proof:} 
Look at the form of $\varrho_{AB}$,
obviously, $\mathcal
{M}=\textbf{0}$ implies $\varrho_{AB}$ is separable. To prove the
converse, one needs the definition of separability:
$\varrho_{AB}=\sum_i p_i\;
 \tau_{Ai}\otimes\tau_{Bi}$,
where $\tau_{Ai}$ and $\tau_{Bi}$ are, respectively, Alice and Bob's
local density matrices, and $p_i>0$ satisfy $\sum_i p_i=1$. For
convenience, let $\tau_{Ai}^{mn} \; (m, n=1, 2)$ denote the element
of Alice's density matrix $\tau_{Ai}$. By calculating ${\rm
tr}_A[(|0\rangle\langle0|\otimes \openone) \varrho_{AB}]$ and ${\rm
tr}_A[(|1\rangle\langle1| \otimes \openone)\varrho_{AB}]$, one has $
\sum_i p_i \;\tau_{Ai}^{11}\tau_{Bi}=\mu_1|\varphi_1\rangle\langle
 \varphi_1|$, $ \sum_i p_i \;\tau_{Ai}^{22}\tau_{Bi}=\mu_2|\varphi_2\rangle\langle
 \varphi_2|.$
Let $|\varphi_1^\perp\rangle$ and $|\varphi_2^\perp\rangle$ be two
pure states that are orthogonal to $|\varphi_1\rangle$ and
$|\varphi_2\rangle$, respectively. Notice that ${\rm tr}[\sum_i p_i
\;\tau_{Ai}^{mm}\tau_{Bi} \times
|\varphi_m^\perp\rangle\langle\varphi_m^\perp|]=0$, $(m=1,2)$,
thus, for any index $i$, we have $\tau_{Ai}^{mm}{\rm
tr}(\tau_{Bi}|\varphi_m^\perp\rangle\langle\varphi_m^\perp|)=0$,
which results in
\begin{eqnarray}
 \tau_{Ai}^{11}\tau_{Ai}^{22}\;
 [\;{\rm tr}(\tau_{Bi}|\varphi_1^\perp\rangle\langle\varphi_1^\perp|)+
 {\rm tr}(\tau_{Bi}|\varphi_2^\perp\rangle\langle\varphi_2^\perp|)\;]=0.
\end{eqnarray}
Since $|\varphi_1^\perp\rangle \neq|\varphi_2^\perp\rangle$, they cannot
be simultaneously perpendicular to the state $\tau_{Bi}$, thus
$\tau_{Ai}^{11}\tau_{Ai}^{22}=0$, which yields $\tau_{Ai}^{12}=\tau_{Ai}^{21}=0$ due to
positivity condition of
$\tau_{Ai}$. So
$ \mathcal {M}=\sum_i p_i\tau_{Ai}^{12}\tau_{Bi}=\textbf{0}$.
Lemma 1 is henceforth proved.


\vspace{8pt}
\noindent\emph{Lemma 2: }The state $\varrho_{AB}$ admits a local-hidden-state (LHS) model (which means that it is not steerable) if and only if $\mathcal {M}=\textbf{0}$.


\noindent\emph{Proof:} $\mathcal {M}=\textbf{0}$ implies
$\varrho_{AB}$ is separable, thus $\varrho_{AB}$ admits a LHS model.
Now we focus on the proof of necessity. If Alice's measurement
setting is $\{\hat{z}, \hat{x}\}$, then one has
\begin{subequations}
\label{E}
\begin{eqnarray}
 \tilde{\rho}^{\hat{x}}_0&=&\frac{1}{2}(\mu_1 |\varphi_1\rangle \langle \varphi_1 |
 + \mu_2 |\varphi_2\rangle \langle \varphi_2| + \mathcal {M}+\mathcal{M}^\dagger),\label{Erhox0}\\
 \tilde{\rho}^{\hat{x}}_1&=&\frac{1}{2}(\mu_1 |\varphi_1\rangle \langle \varphi_1 |
 + \mu_2 |\varphi_2\rangle \langle \varphi_2| - \mathcal {M}-\mathcal {M}^\dagger).\label{Erhox1}
\end{eqnarray}
\end{subequations}
Substitute Eqs.~(\ref{z01a})(\ref{z01b})(\ref{Erhox0})(\ref{Erhox1})
into Eq.~(\ref{E0})
and due to
$\langle\varphi_1^\perp| \tilde{\rho}^{\hat{z}}_0 |\varphi_1^\perp\rangle = 0$ and
$\langle\varphi_2^\perp|\tilde{\rho}^{\hat{z}}_1|\varphi_2^\perp\rangle = 0 $,
one immediately has
$\rho_{\xi} \in \{ |\varphi_1 \rangle \langle\varphi_1 |, |\varphi_2 \rangle
\langle\varphi_2 | \}$ for any $\xi$. Based on which, Eqs.~\eqref{Erhox0} (\ref{Erhox1}) are valid only if
$\mathcal {M}+\mathcal {M}^\dag=(\alpha_x |\varphi_1 \rangle \langle\varphi_1 |
+\beta_x |\varphi_2 \rangle \langle\varphi_2 |)/2$, with $\alpha_x,\beta_x \in \mathbb{R}$.
Similarly, if Alice's measurement setting is $\{\hat{z}, \hat{y}\}$,
then one has $\mathcal {M}-\mathcal{M}^\dag=i(\alpha_y |\varphi_1 \rangle \langle\varphi_1 |
+\beta_y |\varphi_2 \rangle \langle\varphi_2 |)/2$, with $\alpha_y$, $\beta_y\in \mathbb{R}$.
If there exists a LHS model for Bob's states, then
$\mathcal {M} = \alpha |\varphi_1 \rangle \langle\varphi_1 | + \beta|\varphi_2
\rangle \langle\varphi_2 |$, with $\alpha=\alpha_x+i\alpha_y$, $\beta=\beta_x+i\beta_y$.
Substitute $\mathcal {M}$ into Eq.~\eqref{varrhoAB}, we have
\begin{eqnarray}
 \varrho_{AB}=\mu_1 T_\alpha \otimes |\varphi_1\rangle \langle \varphi_1|
+ \mu_2 T_\beta \otimes |\varphi_2\rangle \langle
 \varphi_2 |,\nonumber
\end{eqnarray}
with $T_\alpha=\left(\begin{matrix} 1 & \alpha \\
 \alpha^* & 0 \end{matrix}\right)$ and $T_\beta=\left(\begin{matrix} 0&\beta\\
 \beta^*&1\end{matrix}\right)$. Now we construct the following two projectors: $\mathcal {Q}_1=
 |\chi_1\rangle \langle \chi_1 | \otimes |\varphi_2^\perp \rangle \langle \varphi_2^\perp |$,
$\mathcal {Q}_2= |\chi_2\rangle \langle \chi_2 | \otimes
|\varphi_1^\perp \rangle \langle \varphi_1^\perp |$,
where $|\chi_1\rangle$ is the eigenvector of $T_\alpha$ with eigenvalue $v_1=(1-\sqrt{1+4|\alpha|^2})/2\leq 0$,
and $|\chi_2\rangle$ is the eigenvector of $T_\beta$ with eigenvalue $v_2=(1-\sqrt{1+4|\beta|^2})/2\leq 0$.
Because $\varrho_{AB}$ is a density matrix, one has
\begin{subequations}
\begin{eqnarray}
 {\rm tr}(\varrho_{AB} \mathcal {Q}_1)=v_1 \mu_1 |\langle \varphi_2^\perp |\varphi_1\rangle|^2 \geq0,\nonumber\\
 {\rm tr}(\varrho_{AB} \mathcal {Q}_2)=v_2 \mu_2 |\langle \varphi_1^\perp |\varphi_2\rangle|^2 \geq0.\nonumber
\end{eqnarray}
\end{subequations}
This leads to
$\mathcal {M} = \textbf{0}$. Lemma~2 is henceforth proved.

Three measurement settings were mentioned in the
proof of Lemma~2. This does not mean that we need a three-setting
protocol to show steering. For a given entangled state
$\varrho_{AB}$, a two-setting protocol is enough to demonstrate
steering. Lemma~2 shows that $\mathcal{M}+\mathcal {M}^\dag$ and
$\mathcal {M}-\mathcal {M}^\dag$ cannot be linearly expanded of
$|\varphi_1 \rangle \langle\varphi_1 |$ and $|\varphi_2 \rangle
\langle\varphi_2 |$ simultaneously (because that means $\mathcal
{M}=\textbf{0}$ and $\rho_{AB}$ is separable). For a given
$\varrho_{AB}$, if $\mathcal{M}+\mathcal {M}^\dag \neq (\alpha_x
|\varphi_1 \rangle \langle\varphi_1 |+\beta_x |\varphi_2 \rangle
\langle\varphi_2 |)/2$, then using $\{\hat{z}, \hat{x}\}$ to
demonstrate steering, otherwise using $\{\hat{z}, \hat{y}\}$.



\vspace{12pt} \noindent{\bf Acknowledgements}


J.L.C. is supported by the National Basic Research Program (973
Program) of China under Grant No.\ 2012CB921900 and the NSF of China
(Grant Nos.\ 10975075 and 11175089). A.C. is supported by the
Spanish Project No.\ FIS2011-29400. This work is also partly
supported by the National Research Foundation and the Ministry of
Education, Singapore (Grant No.\ WBS: R-710-000-008-271).


\vspace{8pt} \noindent{\bf Author Contributions}


JLC initiated the idea. JLC, XJY, HYS and CW established the proof.
JLC, CW, AC, LCK and CHO wrote the main manuscript text. HYS and XJY
prepared figures 1 and 2. All authors reviewed the manuscript.

\vspace{8pt} 
\noindent{\bf Supplementary Information} is linked to the online version of
the paper at www.nature.com/nature.


\vspace{8pt} \noindent{\bf Additional Information}


\vspace{8pt}

\noindent\textbf{Competing financial interests:} The authors declare no
competing financial interests.


\vspace{8pt}

\noindent\textbf{Reprints and permission} information is available at
www.nature.com/reprints.


\end{document}